# Adaptation of Whisper models to child speech recognition

*Rishabh Jain*[1], *Andrei Barcovschi*[1], *Mariam Yiwere*[1], *Peter Corcoran*[1], *Horia Cucu*[2]

[1]School of Electrical and Electronics Engineering, University of Galway, Galway, Ireland
[2]Speech and Dialogue Research Laboratory, University Politehnica of Bucharest, Romania

```
rishabh.jain@universityofgalway.ie, a.barcovschi1@universityofgalway.ie,
mariam.yiwere@universityofgalway.ie, peter.corcoran@universityofgalway.ie,
                          horia.cucu@upb.ro
```

## Abstract

Automatic Speech Recognition (ASR) systems often struggle with transcribing child speech due to the lack of large child speech datasets required to accurately train child-friendly ASR models. However, there are huge amounts of annotated adult speech datasets which were used to create multilingual ASR models, such as Whisper. Our work aims to explore whether such models can be adapted to child speech to improve ASR for children. In addition, we compare Whisper child-adaptations with finetuned self-supervised models, such as wav2vec2. We demonstrate that finetuning Whisper on child speech yields significant improvements in ASR performance on child speech, compared to non-finetuned Whisper models. Additionally, utilizing self-supervised Wav2vec2 models that have been finetuned on child speech outperforms Whisper finetuning.

## 1. Introduction

Automatic Speech Recognition (ASR) faces several challenges, including limited training data, untranscribed training data and performance degradation on non-native speech and children's speech. Recent research in ASR tackles some of these problems, especially for adult speech, and therefore ASR on adult speech has reached human-level performance [1]–[4]. However, for child speech, progress has been slow and ASR models still perform poorly. Unlike adult speech data, high quality child speech datasets required for training are limited and challenging to collect and annotate (see the survey in [5]). Additionally, there are inherent differences between adult and child voices in terms of pitch, linguistic and acoustic features, and pronunciation ability [6], [7]. The shorter vocal tract length and higher fundamental frequency [8] of children's voices also add to the complexity of recognizing child speech.

Recent development in self-supervised learning has delivered improvements for child speech. The development of unsupervised pretraining techniques, such as Wav2vec2 [3], has greatly contributed to the progress of child ASR [9]–[11]. However, a finetuning stage on a labeled dataset is required for ASR, which limits their usefulness since finetuning can find patterns within a training dataset and boost performance on the similar datasets but may not generalize to other dataset distributions. The aim of speech recognition systems is to operate with high reliability in diverse environments, without the need for finetuning for the data/deployment distribution of each specific usecase. We reviewed various supervised learning approaches [12]–[14] in child ASR. It was observed that most of these studies included transfer learning approaches from adult to child speech [9], [12], [15], data augmentation methods [16]–[20], or weakly supervised training [14], [15], [21]. Recent findings in supervised learning approaches [22], [23] has demonstrated that pretraining speech recognition models on multiple datasets/domains using supervised methods can enhance the models' robustness and generalization performance on unseen datasets.

In this work, we use a recent State-of-the-Art (SOTA) supervised ASR model, called Whisper. The authors of Whisper [4] have successfully bridged the gap in weakly supervised speech recognition by using large amounts of labeled audio data. They have also broadened the scope of weakly supervised pre-training beyond English-only speech recognition to be multilingual and multitask, showing great performance on different multilingual adult speech datasets [4]. These findings suggest that the scaling of weakly supervised pretraining has been undervalued for speech recognition. We use these Whisper models to provide an analysis of supervised training paradigms on different child speech datasets. We also finetune these models using different combinations of child speech datasets to see the subsequent speech recognition performance on different seen and unseen distributions of child speech datasets [24]–[26]. Lastly, we provide a comparative analysis of Whisper results with previously benchmarked results that used wav2vec2 self-supervised learning approach trained on the same distribution of datasets [27]. We use a similar approach as used by the authors of [28] for providing a comparison between Whisper and wav2vec2 results.

Since Whisper is trained with an order of magnitude more data than wav2vec2 (680k vs 60k) and contains a lot of multilingual and low resource languages during training, we believe that this multilingual data can be utilized to provide child speech recognition tasks via finetuning. Our goal is to evaluate the efficacy of these two methodologies in child speech analysis and determine their potential for enhancing child ASR technology and developing educational tools for children.

## 2. Model Description

### 2.1. **Whisper** [4]

The Whisper approach focuses on broadening the scope of weakly supervised pre-training beyond English-only speech recognition to be both multilingual and multitask. Of the 680,000 hours of labelled audio used by Whisper, 117,000 hours cover 96 other languages. The dataset also includes 125,000 hours of X→en translation data. The model processes audio through a system of transformer blocks with residual connections and final layer normalization. The model uses a multitask format to perform the entire speech processing pipeline, including transcription, translation, voice activity detection, alignment, and language identification. The model is based on an encoder-decoder Transformer, which is fed 80-channel log-Mel spectrograms. The encoder is formed by two convolutional layers with a kernel size of 3, followed by a sinusoidal positional encoding, and a stacked set of Transformer blocks. The decoder uses the learned positional embeddings and the same number of Transformer blocks as the encoder. The Whisper architecture is explained in detail in [4].

### 2.2. Wav2vec2 [3]

Wav2vec 2.0 is a speech recognition model and training approach that is based on a self-supervised learning of speech representations using a two-stage architecture for pretraining and finetuning. The architecture of wav2vec 2.0 can be divided into three main parts: a CNN feature extractor, a transformer-based encoder, and a quantization module (see [3] for more details) . In the pretraining phase, the model is trained on a large dataset of unlabelled speech data. The model learns meaningful representations by capturing the temporal and spectral characteristics of speech using a masked contrastive loss function. In the finetuning phase, the pretrained model is finetuned on a smaller labeled dataset for a specific downstream task. The last layer of the pretrained model is replaced with a task-specific feed-forward layer and the entire model is optimized by minimizing the CTC loss [29] for ASR.

### 2.3. Training details

All models were trained using A6000 GPUs with 48GB of available memory. We provide the architectural parameters details in Table 1 for both Whisper and wav2vec2 models used in this work. Whisper models are trained with a large number of parameters and therefore should provide better generalization towards unseen datasets compared to wav2vec2.

Table 1: *Architecture parameters for Whisper* [4] *and wav2vec2* [3] *models.*

| Models | Layers | Width | Heads | Learning Rate | Para-meters |
|---|---|---|---|---|---|
| **Whisper Models:** | | | | | |
| Tiny | 4 | 384 | 6 | $1.5 \times 10^{-3}$ | 39M |
| Base | 6 | 512 | 8 | $1 \times 10^{-3}$ | 72M |
| Small | 12 | 768 | 12 | $5 \times 10^{-4}$ | 244M |
| Medium | 24 | 1024 | 16 | $2.5 \times 10^{-4}$ | 769M |
| Large | 32 | 1280 | 20 | $1.75 \times 10^{-4}$ | 1550M |
| **Wav2vec2 Models:** | | | | | |
| Base | 12 | 768 | 8 | $5 \times 10^{-4}$ | 95M |
| Large | 24 | 1024 | 16 | $3 \times 10^{-4}$ | 317M |

For finetuning, we use a learning rate of $1 \times 10^{-5}$ for all Whisper finetuning experiments. Wav2vec2-base was finetuned with a learning rate of $1 \times 10^{-4}$, while wav2vec2-large was finetuned with a learning rate of $2.5 \times 10^{-5}$, consistent with [3]. Finetuning both approaches involve training the final layer of the models and freezing all others, as described by the respective authors. Finetuning parameters were kept the same as provided in Whisper [4] and wav2vec2 [3]. The Whisper model undergoes finetuning by minimizing the cross-entropy objective function, whereas wav2vec2 is finetuned by minimizing the CTC loss.

## 3. Corpus Description

The authors of Whisper [4] do not mention the datasets used. However, these trained models achieved SOTA results on many different adult speech ASR datasets [4]. For our work, we use three different child speech datasets and one adult speech dataset: MyST Corpus [24], PFSTAR dataset [25], CMU Kids dataset [26] and LibriTTS dev-clean dataset [30]. The datasets are kept consistent with previous research [27] on wav2vec2 to provide objective comparison with the Whisper models.

### 3.1. Dataset Cleanup

All the labeled data was cleaned as per the guidelines mentioned by the authors of Whisper [4]. The abbreviations, punctuations, white spaces, and other non-alphanumeric characters were removed, and all the characters were changed to lowercase. Audio data was modified to have a 16Khz sampling rate and be 16-bit mono channel. The 'dev-clean' subset of LibriTTS [30], containing 9 hours of audio is used to provide an evaluation of our experiments on adult speech. My Science Tutor (MyST) Corpus [24] is an American English child speech dataset containing over 393 hours of child speech, of which 197 hours are fully transcribed. The dataset was cleaned and prepared as mentioned in [27], with 65 hours of clean child speech divided into two subsets: 55 hours for training and 10 hours of testing. PFSTAR [25] includes a collection of words spoken by British English children and contains a total of 12 hours of audio. 10 hours of this data was used for training and 2 hours was held out for inference. CMU Kids [26] corpus was used for validation-only, which contains 9 hours of read-aloud sentences by children recorded at Carnegie Mellon University. While these may not be very big speech datasets, they currently represent the best publicly available child speech datasets.

### 3.2. Dataset Usage

The datasets were divided according to their usage for 'training' and 'inference'. This information is summarized in Table 2.

Table 2: *Dataset usage*

| Usage | Dataset | Duration |
|---|---|---|
| **Finetuning** (Training) | MyST_55h | 55 hours |
| | PFS_10h | 10 hours |
| **Inference** (Testing) | dev-clean | 9 hours |
| | MyST_test | 10 hours |
| | PFS_test | 2 hours |
| | CMU_test | 9 hours |

## 4. Experiments and Results

### 4.1. Codebase

The Whisper implementation used is provided here [1]. The fairseq [2] implementation of wav2vec2 is used for finetuning experiments. Our trained Whisper models are available to use on the HF platform[3]. The relevant information regarding model training, hyperparameters, graphs/metrics, checkpoints, and dataset availability are made available on our GitHub[4].

---

[1] **Whisper Implementation:** https://github.com/huggingface/community-events/tree/main/whisper-fine-tuning-event
[2] **Wav2vec2 Fairseq:** https://github.com/facebookresearch/fairseq

[3] **Finetuned Whisper models:** https://huggingface.co/rishabhjain16
[4] **GitHub:** https://github.com/C3Imaging/whisper_child_speech

## 4.2. Experiments

In our first set of experiments (see Section 4.3.1), the original Whisper models were evaluated on different child speech datasets mentioned in Table 2. The models are categorized based on their size: Tiny, Base, Small, Medium, Large, and Large V2 (see Table 1). 'Large-V2' was trained for 2.5X more epochs as compared to 'Large', while also adding extra parameters for regularization [4]. There are two versions of each model: one trained with multilingual data and one specifically for the English language only (indicated by '.en' in the name). 'Large' and 'Large-V2' models don't have English-only models. Figure 1 shows a plot comparing Word Error Rate (WER) on 12 English adult speech datasets against model parameters (as provided by Whisper[4]). As expected, lower WER values are obtained using models with more parameters. We also perform a similar comparison using our child speech datasets (more in section 4.3).

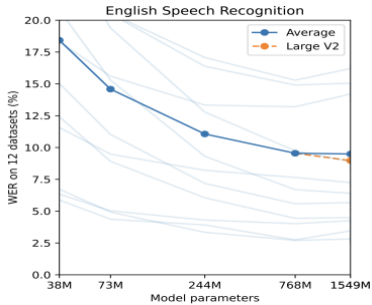

Figure 1: *Whisper Parameters vs. WER on adult speech datasets (from [4]).*

The second set of experiments (see Section 4.3) involved finetuning these Whisper models with child speech. Three models with the best performance from the first set of experiments are selected for further finetuning. We finetuned each of the selected models up to 4000 epochs. We select the best performing checkpoints from among the trained models, which shows the lowest WER while training. Finetuning included three experimental configurations of training data: MyST_55h, PFSTAR_10h, and MyST_55h+PFSTAR_10h combined. These finetuning experiments were kept consistent with previously reported wav2vec2 finetuning experiments [27] in order to compare both models trained with a similar distribution of finetuning data. The wav2vec2 'base' and 'large' models are used for finetuning, which are pretrained with 960 hours of Librispeech data [31], and 60,000 hours of Librilight data [32], respectively. The difference in their parameters sizes can be seen in Table 1. This comparison is provided to see how supervised and self-supervised approaches behave with child speech.

## 4.3. Results and Discussion

### 4.3.1. Whisper Original (No-Finetuning):

Table 3 provides the WER results on the inference datasets using different original Whisper models from the first set of experiments. These models are provided by the authors [4] and no initial finetuning was performed over these models. It can be observed that the models with larger numbers of parameters generally perform better. Among the models with the same number of parameters, the English models perform better than the multilingual models, suggesting that training on language-specific data can improve performance for that language. The lowest WER achieved are highlighted in Table 3.

Table 3: *WER for different Whisper and Wav2vec2 models (without finetuning) on child speech (MyST, PFSTAR and CMU Kids) and adult speech (dev-clean) datasets.*

| Models | MyST_test | PFS_test | CMU_test | dev-clean |
|---|---|---|---|---|
| Tiny | 40.09 | 159.57 | 30.63 | 10.85 |
| Tiny.en | 33.02 | 47.11 | 27.32 | 8.62 |
| Base | 32.14 | 100.07 | 25.03 | 8.14 |
| Base.en | 29.15 | 45.70 | 20.75 | 7.18 |
| Small | 26.22 | 111.75 | 18.52 | 6.43 |
| Small.en | 26.72 | 39.00 | 16.82 | 6.06 |
| Medium | 25.11 | 80.97 | **12.67** | 5.58 |
| Medium.en | 28.06 | **35.25** | 14.00 | 6.20 |
| Large | 25.24 | 84.52 | 13.70 | 5.53 |
| Large-V2 | **25.00** | 73.68 | 12.69 | **5.40** |
| w2v2-base (LS_960) | 15.41 | 11.20 | 16.33 | 3.40 |
| w2v2-large (LL_60k) | **12.50** | **8.56** | 14.85 | **3.28** |

Note: '.en' respresents the English-only trained models, while all others represent the multilingual models. For example, 'Tiny' contains both English and other multilingual training data while 'Tiny.en' contains only English speech. Wav2vec2 results presented for comparison are taken from previously presented work on wav2vec2 for child ASR [27]. The 'w2v2-base' is pretrained with 960 hours of Librispeech data (LS_960) and 'w2v2-large' is pretrained with 60k hours of Librilight data (LL_60k). Both models were finetuned using Librispeech for providing a comparison with non-finetuned Whisper models. The WER reported in Table 3 uses zero-shot setting.

These models achieved positive results on multilingual adult speech without the need to perform data-specific finetuning (see Figure 1), however, the performance seems poor for child speech, despite Whisper stating that their models generalize well to standard benchmarks in a zero-shot transfer setting without the need for any finetuning. We use these experiments as a baseline for further finetuning. The models with lowest WER were chosen ('Medium', 'Medium.en' and 'Large-V2') for providing further finetuning with child speech.

### 4.3.2. Whisper Finetuning with Child Speech

The Whisper finetuning experiments include three subsets of experiments: finetuning with MyST_55h, PFSTAR_10h and a combination of both datasets. Table 4 shows the WER of the selected finetuned models using these subsets. During finetuning, cross entropy loss is minimized by training only on the last layer and freezing all other layers, allowing the model to classify target tokens from a predefined vocabulary.

Table 4: *WER on inference (test) datasets for different Whisper and wav2vec2 models finetuned on MyST, PFSTAR and MyST+PFSTAR-combined datasets.*

| ID | Models | MyST_test | PFS_test | CMU_test | dev-clean |
|---|---|---|---|---|---|
| **MyST (55 Hours) Finetuning:** | | | | | |
| 1 | Medium | **11.66** | 19.76 | 16.84 | 5.62 |
| 2 | Medium.en | 11.81 | 17.83 | **15.07** | 6.48 |
| 3 | Large-V2 | 12.28 | 10.88 | 15.67 | **4.82** |
| 4 | w2v2-base | 8.13 | 14.77 | 16.47 | 7.72 |
| 5 | w2v2-large | **7.51** | 12.46 | 15.25 | 6.43 |
| **PFSTAR (10 Hours) Finetuning:** | | | | | |
| 6 | Medium | 16.18 | 3.15 | 16.57 | 5.33 |
| 7 | Medium.en | 15.84 | 3.14 | 15.53 | 5.28 |
| 8 | Large-V2 | **15.79** | 2.88 | 15.22 | **5.10** |
| 9 | w2v2-base | 31.86 | **3.48** | 27.49 | 13.95 |
| 10 | w2v2-large | 27.17 | 3.50 | 21.35 | 11.60 |
| **MyST (55 Hours) + PFSTAR (10 Hours) Finetuning:** | | | | | |
| 11 | Medium | **12.22** | 2.98 | 16.05 | 5.40 |
| 12 | Medium.en | 12.33 | 3.32 | **15.08** | 4.88 |
| 13 | Large-V2 | 13.34 | 4.17 | 17.11 | 4.97 |
| 14 | w2v2-base | 7.94 | **2.91** | 15.97 | 7.64 |
| 15 | w2v2-large | **7.42** | 2.99 | 14.18 | 5.79 |

Note: Wav2vec2 results are taken from [27]. The 'w2v2-base' represents wav2vec2 base model while 'w2v2-large' represents wav2vec2 large models.

Finetuning with MyST_55h showed a significant improvement in the WER of MyST_test and PFS_test. However, CMU_test dataset had a 2% increase in WER, as shown in Table 4. WER on dev-clean adult speech dataset also decreased by 1%. Finetuning with PFS_10h also had a significant improvement on MyST_test and PFS_test. The WER on both test sets decreased; however, the improvement in WER on the MyST_test is not as good as when the models are finetuned with MyST_55h. CMU_test had a 2% increase in WER, similar to MyST_finetuning. Large-V2 Whisper model gave the lowest WER on all four inference data setups, with WER on PFS_test dropping to 2.88. When both MyST_55h and PFS_10h were used for finetuning, the WER on both MyST_test and PFS_test dropped significantly. It can be observed that for a dataset used in finetuning, the model shows an improvement in performance on datasets with similar distribution at inference time.

The following observations were seen in all finetuning experiments: Whisper finetuned models yield better results than Whisper original models, regardless of dataset distribution, but a finetuning dataset that matches the distribution of the test dataset can improve performance. CMU_test showed an increase in WER regardless of the finetuning setup and remained in the range of 15-17%. This could imply that CMU Kids might be a noisy dataset which doesn't work well for ASR. The WER of dev-clean adult speech further decreased after child speech finetuning and stayed in the range of 4-5% for all experiments.

*4.3.3. Whisper vs Wav2vec2:*

We compare Whisper models with wav2vec2 finetuned models on the same datasets. Table 3 and Table 4 cover the various wav2vec2 finetuning results on different child speech datasets. We first compare Librispeech-finetuned 'base' and 'large' wav2vec2 models with the original Whisper 'Medium' and 'Large' models (See Table 3). This was done to maintain consistency with the comparison mechanism as provided by authors of Whisper [4]. The wav2vec2 models finetuned with Librispeech generally performed better on child speech compared to any of the Whisper models without finetuning. Both these models were used to provide a usecase of ASR over unseen child speech in low resource data scenario. Wav2vec2 results show the lowest WER on all inference datasets except CMU_test. However, Whisper models gave lower WER on CMU_test as compared to wav2vec2 models. This implies that CMU kids dataset could have acoustic properties similar to adult speech since supervised finetuning using Whisper decreases the WER on CMU_test.

The results of the experiments with child speech finetunings show that wav2vec2 finetuning using MyST_55h resulted in lower WER compared to Whisper finetuning on MyST_test. However, an increase in WER was observed on PFS_test and dev-clean for wav2vec2 finetuning. Both Whisper and wav2vec2 finetuned models had a WER range of 14-16% on CMU_test. For PFS_10h finetuning, similar results were obtained for both wav2vec2 and Whisper models on PFS_test, with WER of 3.48 and 2.88, respectively. However, high WERs were observed on all other inference datasets. These results suggest that wav2vec2 finetuning generalizes well for datasets with a similar distribution, while Whisper finetuning works best for unseen datasets at inference time. When both MyST_55h and PFS_10h were used for finetuning, the lowest WER was observed with wav2vec2 finetuning across all child speech datasets as compared to Whisper finetuning. Both Whisper and wav2vec2 models behaved similarly when finetuned with a combination of child speech datasets, but wav2vec2 performed better on datasets with similar distributions as the seen datasets. Moreover, when considering the amount of training data and model size (model 13 vs model 14), it was observed that the wav2vec2 model 15 (60k hours, 317M parameters) performed better than Whisper model 13 (680k hours, 1550M parameters), which were finetuned with the same amount of child speech data. These findings demonstrate that wav2vec2 performs well with child speech and slightly outperforms Whisper.

## 5. Conclusions

In this paper, we use the recent SOTA large-scale supervised Whisper models for experimental analysis over different child speech datasets. The study of different combinations of finetuning over child-specific datasets is also presented in this paper. Finetuning Whisper models achieved significant improvements in accuracy of child speech recognition. We also present comparisons with the SOTA self-supervised, wav2vec2 model. Finetuning both Whisper and wav2vec2 improves performance of child ASR. While Whisper improves ASR performance for both adult and child speech, regardless of the finetuning dataset, wav2vec2 model performs better with finetune-specific datasets. Although Whisper may be more appropriate for unseen datasets, wav2vec2 is a better choice for real-time, task-specific applications. In addition, the use of smaller-sized models, such as wav2vec2, would be more feasible for deployment on edge devices, which is also using 10 times less training data than Whisper. For future work, we aim to further study this methodology by including more low resource datasets (both adult and child), different ASR decoding strategies and deploying these models on edge devices.

## 6. Acknowledgements

The authors would like to acknowledge experts from Xperi Ireland: Gabriel Costache, Zoran Fejzo, and George Sterpu for providing their expertise and feedback while working on this research.